\documentclass[twocolumn,showpacs,preprintnumbers,superscriptaddress,amsmath,amssymb]{revtex4}
\usepackage{graphicx}
\usepackage{dcolumn}
\usepackage{bm}
\input psfig.sty

\newcommand{\gsim}{\mathrel{\vcenter{\hbox{$>$}\nointerlineskip\hbox{$\sim$}}}}

\newcommand{\su}{\textsc{s} \hspace{-0.34em} \textrm{/}}

\begin{document}

\title{\bf Casimir Effect in a Supersymmetry-Breaking Brane-World as Dark Energy} 

\author{Pisin Chen} %
\affiliation{Stanford Linear Accelerator Center, Stanford
University, Stanford, CA 94309, USA}
\author{Je-An Gu} %
\affiliation{Department of Physics, National Taiwan University,
Taipei 106, Taiwan, R.O.C.} %
\affiliation{Stanford Linear Accelerator Center, Stanford
University, Stanford, CA 94309, USA}

\date{\today}

\begin{abstract}
A new model for the origin of dark energy is proposed based on the
Casimir effect in a supersymmetry-breaking brane-world.
Supersymmetry is assumed to be preserved in the bulk while broken
on a 3-brane. Due to the boundary conditions imposed on the
compactified extra dimensions, there is an effective Casimir
energy induced on the brane. The net Casimir energy contributed
from the graviton and the gravitino modes as a result of
supersymmetry-breaking on the brane is identified as the observed
dark energy, which in our construction is a cosmological constant.
We show that the smallness of the cosmological constant, which
results from the huge contrast in the extra-dimensional volumes
between that associated with the 3-brane and that of the bulk, is
attainable under very relaxed conditions.
\end{abstract}



\maketitle


Recent type Ia supernova (SN Ia) observations suggest that
the expansion rate of the universe is increasing 
\cite{Perlmutter:1999np,Riess:1998cb}. Possible explanations of
this acceleration include a form of energy which provides a
significant negative pressure, such as a positive cosmological
constant \cite{Lambda models}, a positive dynamical potential
energy induced by a scalar field called quintessence
\cite{Caldwell:1998ii,ComplexQ}, a string-theory-induced
metastable de Sitter vacuum \cite{Kachru:2003aw}, etc. There are
also ideas that invoke the existence of extra spatial dimensions
\cite{Dvali:2002,Gu:DE-ED,Burgess:2003&2004} and the modification
of gravity \cite{Carroll:2003wy,Lue:2003ky,Arkani-Hamed:2003uy}.

On the experimental side, the cosmological constant has survived
through an array of updated high-precision observations.
Unfortunately, on the theoretical side it suffers a generic
fine-tuning problem common to most attempts. In particular, the
idea faces a severe challenge in that the contribution of quantum
fluctuations to the vacuum energy, and thus the value of the
cosmological constant, is generally much larger than what is
suggested by observation. This is the long-standing cosmological
constant problem.

Before the discovery of the present accelerating expansion of the
universe in 1998, the cosmological constant problem was how to
make it vanish (pre-dark-energy). After 1998, there came another
problem
--- how to make the cosmological constant slightly deviate from zero
(post-dark-energy). The pre-dark-energy problem stems from our
lack of an ultimate understanding of accommodating quantum vacuum
in gravity. While it still awaits a profound answer from a future
theory that would successfully combine general relativity and
quantum theory, as we shall review below, some interesting
proposals are already in sight.
The post-dark-energy problem, on the other hand, appears soluble
based on our current knowledge. 

One interesting early idea for ameliorating the pre-dark-energy
cosmological constant problem is, instead of making it small, to
invoke extra dimensions such that the expansion rate of the
imbedded (3+1) space-time is independent of the vacuum energy
\cite{Rubakov:1983}. This idea has received reviving interests in
the post-dark-energy era, where physical models have been proposed
based on the brane-world scenario. One such approach, for example,
involves a codimension-two brane whose brane tension induces a
deficit angle in the bulk, which in turn cancels the brane tension
exactly \cite{Chen:2000at,Burgess:2003&2004,Vinet:2004}. Another
approach \cite{Arkani-Hamed:2000eg,Kachru:2000,Carroll:2001zy} is
to construct a modified Friedmann equation with the form $H^2
\propto (\rho + p)$, so that the vacuum energy (with the equation
of state $\rho = -p$) would not contribute to the Hubble
expansion. We shall refer to this generic idea, including the two
types of models described above, as the self-tuning mechanism.

Even if the pre-dark-energy cosmological constant problem may
eventually be solved by either the self-tuning mechanism or some
as-yet undiscovered novel concept, the post-dark-energy problem
would still remain. The tremendous hierarchy between the vacuum
energy implied by the observations and the known particle physics
energy scales remains a severe theoretical challenge by itself.
What underlying physics could be responsible for such a tiny
vacuum energy? In this Letter we focus on this latter issue,
assuming that the pre-dark-energy problem can eventually be
addressed. We point out that the same ingredients invoked in the
self-tuning mechanism, that is, the extra dimensions and the
brane-world scenario, when augmented with supersymmetry (SUSY),
can in principle solve the post-dark-energy problem with very
relaxed constraints.

Specifically, we consider the Casimir effect as the origin for the
dark energy. It is known that positive Casimir energy in the
ordinary $(3+1)$-dimensional space-time cannot provide negative
pressure. Conversely, the Casimir energy induced from a
higher-dimensional world with suitable boundary conditions in
extra dimensions can, in principle, behave like a cosmological
constant in the imbedded 3-space. Such a Casimir energy
nevertheless tends to be too large for this purpose unless the
size of extra dimensions is macroscopic, as summarized by Milton
\cite{Milton:2002hx}. We show that such a limitation can be
ameliorated if one further invokes SUSY in the system.

SUSY guarantees the perfect cancellation of the vacuum energy.
Unfortunately (or fortunately, as we will see), we also know that
SUSY has to be broken, at least in our (3+1)-dimensional world,
with the symmetry-breaking scale $\gsim$ TeV. Conventionally this
would entail a vacuum energy that is much too large for dark
energy. But as we have assumed, the brane tension so induced is to
be cancelled perfectly by the self-tuning mechanism or its
variant. If we further assume that SUSY is broken on the brane
through, for example, a gauge-mediated SUSY-breaking mechanism
(for a review, see \cite{Giudice:1998bp}) while preserved in the
bulk, then in this configuration we find that the leading
contribution to the vacuum energy a la Casimir effect can be
dramatically suppressed relative to the SUSY-breaking scale.

We emphasize that there is a fundamental difference between such a
Casimir energy and the conventional vacuum energy. The Casimir
energy in our SUSY configuration is nontrivial only around the
3-brane, and, in particular, it entails the equations of state:
$p_a=-\rho$ and $p_b
> 0$, where $p_a$ and $p_b$ are its pressures along the 3-brane
and the extra dimensions, respectively. In contrast, the brane
tension from the conventional vacuum energy obeys the following
equations of state: $p_a=-\rho$ and $p_b = 0$, which are a
necessary condition for its cancellation via the self-tuning
mechanism. Thus the Casimir energy cannot be removed by the same
self-tuning procedure and should survive as the leading
contribution to vacuum energy on the brane.

A similar concept of SUSY-breaking in a brane-world, called
supersymmetric large extra dimensions (SLED), has been invoked to
address the vacuum energy and the cosmological constant problem
\cite{Burgess:2003&2004}. In this proposal SUSY in the bulk is
slightly broken by the presence of non-supersymmetric 3-branes. By
incorporating the self-tuning mechanism for two extra dimensions,
the brane tension is exactly cancelled with the curvature
\cite{Chen:2000at}, and as a result the leading contribution to
vacuum energy is induced by the weak SUSY-breaking in the bulk.
The smallness of dark energy in this proposal relies on the
requirement of two large extra dimensions of the size around
$0.1\,$mm, which is on the edge of the current experimental
constraint \cite{Long:2003dx}.

In our approach the smallness of the dark energy is attained
through the very large contrast between the volume of the
SUSY-breaking region around the brane and that of the
SUSY-preserving region in the bulk, where their ratio naturally
arises when one integrates out the imbedding extra dimensions to
obtain the Casimir energy on the 3-brane. To demonstrate the
powerfulness of this new way of handling the post-dark-energy
problem, we examine such a Casimir effect under a wide range of
extra dimensions, without limiting ourselves to the specific 1 and
2 extra dimensions invoked in the self-tuning models.

Let us consider a (3+$n$+1)-dimensional space-time with $n$
compact extra dimensions, in which the standard model fields and
their superpartners are confined on a 3-brane while the gravity
(graviton-gravitino) sector  resides 
in the (higher-dimensional) bulk. We assume SUSY is preserved in
the bulk and only broken around the 3-brane with a breaking scale
$M_{\textsc{susy}}$. Unbroken SUSY dictates that gravitons and
gravitinos in the bulk have the same mass and the same interaction
strength, while its breakage around the 3-brane, in our
assumption, induces a mass-square difference $\mu^2$ between them.
The SUSY-breaking region around the 3-brane in general has a
nonzero effective thickness $\delta$ that, motivated by string
theory, is characterized by the string length $l_\textsc{s}$.
Consequently, the vacuum energy as well as the Casimir energy
vanish in the bulk but is nontrivial in the extra-dimensional
volume that encompasses the brane with thickness $\delta$.

In general, the evolution and the curvature of the extra
dimensions, in addition to the energy sources in the bulk and on
the brane, can affect the evolution of the universe. In many
recent brane-scenario-based models extra dimensions are made
stable around a specific size by a certain mechanism (e.g., see
\cite{Arkani-Hamed:1998kx}) so as to
satisfy the experimental constraints from the solar-system tests
of gravity. For simplicity we consider the stabilization of the
extra dimensions via a toroidal geometry (i.e.\ with zero
curvature), so that the extra dimensions provide no effect on the
evolution of our 3-dimensional world. We can therefore employ the
ordinary (3+1)-dimensional Einstein equations in which the
energy-momentum tensor is obtained from that of the imbedding
higher-dimensional space-time through the integration over the
extra space.

As a demonstration of how the mass shift of bulk fields induced by
SUSY-breaking around the brane modifies
the Casimir energy, we first consider the case of a     
scalar field and its superpartner, ``{\it calar}'', in a
Minkowskian ordinary (3+1)-dimensional space-time
($\mathcal{M}^4$) and a n-torus extra space ($\mathcal{T}^n$) with
a size $a$. Later we will generalize this derivation to the true
graviton-gravitino case and to the situation where there are more
SUSY fields in the bulk.

Let us employ the form $\Delta m^2(y) = m^2 e^{-2|y|/\delta}$ to
characterize the mass-square shift, where $y$ is the
extra-dimension coordinate and we have set the location of the
3-brane at $y=0$. With the scalar-field part of the action
\begin{equation}
S = \int d^4 x d^n y \sqrt{|g|} \left[ \frac{1}{2} \left(
\partial \phi \right)^2 - \frac{1}{2} \left( m_0^2 + \Delta m^2
\right) \phi^2 \right] ,
\end{equation}
we treat the mass-square-shift term $\Delta m^2 \phi^2$ as a
perturbation and calculate the Casimir energy shift to the first
order.

In $\mathcal{M}^4 \times \mathcal{T}^n$ the renormalized Casimir
energy density $\rho_{\textrm{v}}^{\textrm{(ren)}}(m^2,a)$ is the
difference of the vacuum energy densities at $a$ and infinity:
\begin{equation}
\rho_{\textrm{v}}^{\textrm{(ren)}}(m^2,a) = \rho_{\textrm{v}}
(m^2,a) - \rho_{\textrm{v}} (m^2,a \rightarrow \infty) \, .
\end{equation}
Up to $\mathcal{O}\left(\Delta m^2\phi^2\right)$, the shift of the
Casimir energy density due to SUSY-breaking is
\begin{eqnarray}
\delta \rho_{\textrm{v}}(\Delta m^2,a) & \equiv &
\rho_{\textrm{v}}^{\textrm{(ren)}}(m^2,a) -
\rho_{\textrm{v}}^{\textrm{(ren)}} (m^2=0,a) \\
& \cong & C_n (m_0 a) \cdot \frac{a^2}{a^{4+n}} \cdot  \Delta
m^2(y) \, .
\end{eqnarray}
The coefficient $C_n$ is in general strongly suppressed for large
$m_0 a$. We thus focus on the case where $m_0 a \ll 1$ and in this
limit $C_n$ is approximately independent of $m_0 a$. Note that the
graviton-gravitino sector corresponds to $m_0 = 0$. As a
demonstration of how $C_n$ varies under different boundary
conditions 
and dimensionalities, selected values of $C_n$ for a real scalar
field
are listed in Table \ref{Table 1}, where the cases for the
periodic (PBC) and the anti-periodic (APBC) boundary conditions
for $\mathcal{M}^4 \times \mathcal{T}^n$ are investigated.

\begin{table}[h!]
\caption{\label{Table 1} Selected values of $C_n$ for a real
scalar field$^{*}$}
\begin{tabular}{|ccc|ccc|ccc|} \hline %
$\; \, n \; \,$ & $\mathcal{T}^n_\textsc{pbc}$ &
$\mathcal{T}^n_\textsc{apbc}$ & $\; \, n \, \;$ &
$\mathcal{T}^n_\textsc{pbc}$ & $\mathcal{T}^n_\textsc{apbc}$ & $\;
\, n \, \;$ & $\mathcal{T}^n_\textsc{pbc}$ &
$\mathcal{T}^n_\textsc{apbc}$
\\ \hline
 1 & $0.015$ & $-0.011$ &  9 & $0.10 $ & $-0.061$ & 17 & $3.2  $ & $-3.0  $ \\ %
 2 & $0.024$ & $-0.012$ & 10 & $0.14 $ & $-0.089$ & 18 & $5.7  $ & $-5.5  $ \\ %
 3 & $0.031$ & $-0.013$ & 11 & $0.19 $ & $-0.14 $ & 19 & $10   $ & $-10   $ \\ %
 4 & $0.038$ & $-0.015$ & 12 & $0.27 $ & $-0.21 $ & 20 & $20   $ & $-19   $ \\ %
 5 & $0.045$ & $-0.018$ & 13 & $0.41 $ & $-0.34 $ & 21 & $38   $ & $-38   $ \\ %
 6 & $0.054$ & $-0.023$ & 14 & $0.65 $ & $-0.57 $ & 22 & $76   $ & $-75   $ \\ %
 7 & $0.065$ & $-0.031$ & 15 & $1.1  $ & $-0.96 $ &    &         &          \\ %
 8 & $0.080$ & $-0.043$ & 16 & $1.8  $ & $-1.7  $ &    &         &          \\ %
\hline
\end{tabular}
\vspace{-0.5em} %
\footnotesize \flushleft{\hspace{1em} $^{*}$ For
a complex scalar field, the value of $C_n$ for each case \\
\hspace*{1.8em} is doubled.}
\end{table}

When the extra-dimensional space is integrated over, the Casimir
energy density in $\mathcal{M}^4$ is reduced to
\begin{eqnarray}
\delta \rho_{\textrm{v}}^{\textrm{(4)}} %
&\cong& C_n \cdot \frac{1}{a^{4}} \cdot m^2 \, a^2 \cdot
\frac{\pi^{n/2} \Gamma(n)}{2^{n-1} \Gamma(n/2)} \left(
\frac{\delta}{a} \right)^n  \\
&=& C_n \cdot \frac{1}{a^{4}} \cdot m^2 \, a^2 \cdot n! %
\left( \frac{V_{\delta}}{V_a} \right) ,
\end{eqnarray}
where $V_{\delta}$ is the volume of the extra-dimensional space
inside which SUSY is broken and $V_a$ is the total volume of the
extra space. We see that if $V_{\delta} \ll V_a \,$, the ratio of
these two volumes would provide a powerful suppression to the
Casimir energy.

Dictated by the nature of SUSY, the Casimir energy density (shift)
of a superpartner has the same functional form (and therefore the
same constant $C_n$), but with an opposite sign. So the above
calculations can be directly applied to the calar field. That is,
$\delta \rho_\textrm{calar} (\Delta \widetilde{m}^2 , a) = -
\delta \rho_\textrm{scalar} (\Delta \widetilde{m}^2 , a)$.
Consequently the net Casimir energy density contributed from the
scalar-calar system induced by SUSY-breaking is
\begin{eqnarray}
\delta \rho_{\su} & = & \delta \rho_\textrm{scalar} (\Delta m^2 ,
a) + \delta \rho_\textrm{calar} (\Delta
\widetilde{m}^2 , a)  \\
& = & \delta \rho_\textrm{scalar} (\Delta m^2, a) - \delta
\rho_\textrm{scalar} (\Delta \widetilde{m}^2, a) ,
\end{eqnarray}
which, up to the first order, is equal to $\delta
\rho_\textrm{scalar} (\mu^2, a)$, where $\mu^2 \equiv \Delta m^2 -
\Delta \widetilde{m}^2$. We note that among possible geometries
and boundary conditions there exist ample choices where $C_n \mu^2
> 0$, such that the positivity of the resultant Casimir energy is
ensured.

We are actually more interested in the Casimir energy density
induced by the gravity sector under SUSY-breaking, where the
gravitino is assumed to have acquired a mass shift around the
brane. To apply the above results to the graviton-gravitino case
in the bulk, which possesses more degrees of freedom than that in
the scalar-calar case, we introduce a numerical factor $N$ to
account for their extra contributions. So for the
graviton-gravitino system we have
\begin{equation}
\delta \rho_{\su}^{\textrm{(4)}} \cong N
C_n \cdot \frac{1}{a^{4}} \cdot \mu^2 \, a^2 \cdot n! %
\left( \frac{V_{\delta}}{V_a} \right) .
\end{equation}
Generally there may be more than one pair of SUSY fields in the
bulk. In that case the total Casimir energy density becomes
\begin{eqnarray}
\delta \rho_{\su,\textrm{total}}^{\textrm{(4)}} & \cong & \left(
\sum_{i} N_{i} C_{n} \mu_{i}^2 \right)
\cdot \frac{1}{a^{2}} \cdot n! \left( \frac{V_{\delta}}{V_a} \right) %
\\
& = & \alpha_n \cdot \mu_{\textsc{max}}^2 \delta^n a^{-(n+2)} \, , \label{rho4} \\
\alpha_n &\equiv& \left( \sum_{i} N_{i} C_{n} \frac{\mu_{i}^2}{\mu_{\textsc{max}}^2} \right) %
\cdot \frac{\pi^{n/2} \Gamma(n)}{2^{n-1} \Gamma(n/2)} \, ,
\end{eqnarray}
where the subscript `$i$' denotes the $i$-th SUSY field and
$\mu_{\textsc{max}}^2$ denotes the largest mass-square shift among
those SUSY fields in the bulk. As a result, the SUSY fields which
possess overwhelmingly large mass-square shifts dominate the
Casimir energy.

By insisting $\rho_{\su,\textrm{total}}^{\textrm{(4)}}$ as dark
energy, we are in effect imposing a constraint on several relevant
fundamental physical quantities. Using the relation between the
Planck scale $M_\textrm{pl}$, the fundamental gravity scale
$M_\textsc{g}$, and the extra-dimension size $a$, $M_\textrm{pl}^2
= M_\textsc{g}^{n+2} a^n $, assuming $\delta \sim l_{\textsc{s}} =
M_{\textsc{s}}^{-1} $, and introducing the ratio $\eta$ of the
(dominant) mass shift $\mu_\textsc{max}$ to the SUSY-breaking
scale $M_\textsc{susy}$, we rewrite Eq.\ (\ref{rho4}) as
\begin{equation}
\delta \rho_{\su,\textrm{total}}^{\textrm{(4)}} \sim \alpha_n
\eta^2 M_\textsc{susy}^2 M_{\textsc{s}}^{-n}
M_\textrm{pl}^{-2(n+2)/n} M_\textsc{g}^{(n+2)^2/n} .
\end{equation}
We further identify the Casimir energy as the dark energy with the
density $\sim 3 \times 10^{-11}\,$eV$^4$. We then arrive at the
following constraint among the several energy scales:
\begin{equation}
\left( \frac{M_\textsc{s}}{M_\textrm{pl}} \right)^{-n} \left(
\frac{M_\textsc{g}}{M_\textrm{pl}} \right)^{(n+2)^2/n} \left(
\frac{ M_\textsc{susy}}{M_\textrm{pl}} \right)^2 \sim 10^{123}
\cdot
\alpha_n^{-1} \eta^{-2} . %
\label{M relation}
\end{equation}
This constraint is quite loose, i.e., it can be satisfied by a
wide range of $M_\textsc{s}$, $M_\textsc{g}$ and
$M_\textsc{susy}$. Its looseness indicates that the smallness of
the dark energy can be easily achieved in our model.  In the
following we will see that this constraint remains flexible even
after additional conditions are imposed.

Although there is no a priori reason why these scales should be
related, it is desirable to reduce the large hierarchy among
various energy scales. With this in mind, we impose further
conditions in Eq.\ (\ref{M relation}): (a) $M_\textsc{susy} \sim
M_\textsc{g}$, i.e.\ bridging the hierarchy between the
SUSY-breaking scale and the fundamental gravity scale; (b)
$M_\textsc{s} \sim M_\textsc{g} \sim M_\textsc{susy}$, i.e.\
insisting that there is only one energy scale in our physics.

\begin{figure}[h]
\centerline{\psfig{file=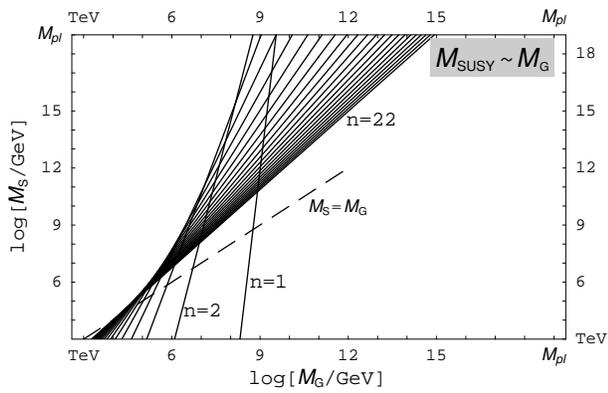,width=8cm}} %
\caption{\footnotesize{\label{Fig1}Constraint on $M_\textsc{s}$
and $M_\textsc{g}$ under the assumption of gravitino dominance:
$\mu_\textsc{max} \sim M_\textsc{susy}^2 / M_\textrm{pl}$ (i.e.\
$\eta \sim M_\textsc{susy} / M_\textrm{pl}$). The solid curves
correspond to solutions under the further assumption of
$M_\textsc{susy}=M_\textsc{g}$ and the dashed line indicates the
condition $M_\textsc{s}=M_\textsc{g}$}}
\end{figure}

Let us focus on the scenario where the mass shift is dominated by
that of the gravitino, which is suppressed by the Planck scale:
$\mu \sim M_\textsc{susy}^2 / M_\textrm{pl}$ (i.e.\ $\eta \sim
M_\textsc{susy} / M_\textrm{pl}$). We further assume that the
values of $\alpha_n$ do not vary too drastically. Then in case (a)
our general constraint, Eq.\ (\ref{M relation}), is reduced to a
more specific constraint on $M_\textsc{s}$ and $M_\textsc{g}$
under different choices of the extra-dimensionality, as
represented by solid curves in Fig.\ \ref{Fig1}. If we further
insist on condition (b), then the solutions further reduce to the
intersects between the line for $M_\textsc{s}=M_\textsc{g}$ and
the solid curves. We find that in case (a) $M_\textsc{g}$ cannot
exceed $10^{15}\,$GeV while the string scale $M_\textsc{s}$ is
barely restricted. In case (b), the specified value of these
quantities is restricted in the range between TeV and
$10^{9}\,$GeV. We note that for larger $n$'s, it approaches TeV, a
soon-to-be testable scale.


There exist various hierarchy problems in physics, such as the
weakness of gravity and the smallness of the cosmological
constant. To reveal the fundamental laws of nature, it is often
desirable to relate the origins of the hierarchies to more
profound physics. In our model, the smallness of the cosmological
constant is a manifestation of the sharp contrast between the
volume of the SUSY-breaking region around the brane and that of
the SUSY-preserving bulk. Note that by further invoking
$M_\textsc{susy} \sim M_\textsc{g}$, our model manages to solve
both hierarchy problems at once. Since our model is based on the
similar ingredients as those in the self-tuning mechanism, it
provides much hope that a synergy between these two concepts may
be found for an eventual complete solution to both pre- and
post-dark-energy problems.


We thank Keshav Dasgupta, Michael Peskin and Marina Shmakova for
useful discussion and comments. This work is supported by the US
Department of Energy under Contract No.\ DE-AC03-76SF00515, and in
part by the Ministry of Education (MoE 89-N-FA01-1-4) and the
National Science Council (NSC 93-2811-M-002-033; NSC
93-2112-M-002-047) of Taiwan, R.O.C. \ JAG wishes to thank SLAC
for its support and hospitality for his visit, during which this
research was carried out.

\end{document}